\documentclass[aps,prl,twocolumn,groupedaddress,notitlepage,showpacs,floatfix,superscriptaddress]{revtex4-1}
\pdfoutput=1
\usepackage{graphicx,graphics,epsfig,subfigure,times,bm,bbm,amssymb,amsmath,amsthm,mathrsfs,MnSymbol}
\usepackage{gensymb}
\usepackage{amsfonts}
\usepackage{float}
\usepackage[matrix,frame,arrow]{xypic}
\usepackage[pdfstartview=FitH]{hyperref}
\hypersetup{
    colorlinks=true,       
    linkcolor=red,          
    citecolor=magenta,        
    filecolor=magenta,      
    urlcolor=cyan,           
    runcolor=cyan
}
\usepackage[pdftex]{color}
\usepackage{hypernat}
\usepackage{braket}
\usepackage{enumerate}
\usepackage[normalem]{ulem}
\usepackage[usenames,dvipsnames]{xcolor}
\usepackage{multirow}
\usepackage{mathtools}
\usepackage{epstopdf}

\definecolor{orange}{rgb}{1,0.5,0}

\newcommand{\ignore}[1]{}
\usepackage{geometry}

\ignore{
\documentclass[eprintnumbers,amsmath,amssymb,onecolumn,a4paper,caption ]{article}
\usepackage{amsfonts}
\usepackage{amssymb}
\usepackage{mathrsfs}
\usepackage{mathbbold}
\usepackage{bbm}
\usepackage{mathrsfs}
\usepackage{dcolumn}
\usepackage{bm}
\usepackage{times,epsfig,amssymb,amsmath}
\usepackage{float}
\usepackage{subfigure}
\usepackage{geometry}\geometry{left=2.5cm,right=2.5cm,top=3cm,bottom=3cm}

\usepackage{color}

}

\begin{document}
	\title{A Universal Quantum Computing Virtual Machine}
	
	\author{Qian-Tan Hong}
	\affiliation{Institute of Physics, Chinese Academy of Sciences, Beijing 100190, China}
	\affiliation{School of Physics, Peking University, Beijing 100871, China}
	
	\author{Zi-Yong Ge}
	\affiliation{Institute of Physics, Chinese Academy of Sciences, Beijing 100190, China}
	\affiliation{School of Physical Sciences, University of Chinese Academy of Sciences, Beijing 100190, China}
	
	\author{Wen Wang}
	\affiliation{Institute of Physics, Chinese Academy of Sciences, Beijing 100190, China}
	\affiliation{School of Physics, Peking University, Beijing 100871, China}
	
	\author{Hai-Feng Lang}
	\affiliation{Institute of Physics, Chinese Academy of Sciences, Beijing 100190, China}

	\author{Zheng-An Wang}
	\affiliation{Institute of Physics, Chinese Academy of Sciences, Beijing 100190, China}
	\affiliation{School of Physical Sciences, University of Chinese Academy of Sciences, Beijing 100190, China}
	
	\author{Yi Peng}
	\affiliation{Institute of Physics, Chinese Academy of Sciences, Beijing 100190, China}
	\affiliation{School of Physical Sciences, University of Chinese Academy of Sciences, Beijing 100190, China}
	
	\author{Jin-Jun Chen}
	\affiliation{Institute of Physics, Chinese Academy of Sciences, Beijing 100190, China}
	\affiliation{School of Physical Sciences, University of Chinese Academy of Sciences, Beijing 100190, China}
	
	\author{Li-Hang Ren}
	\affiliation{Institute of Physics, Chinese Academy of Sciences, Beijing 100190, China}
	\affiliation{School of Physical Sciences, University of Chinese Academy of Sciences, Beijing 100190, China}
	
	\author{Yu Zeng}
	\affiliation{Institute of Physics, Chinese Academy of Sciences, Beijing 100190, China}
	\affiliation{School of Physical Sciences, University of Chinese Academy of Sciences, Beijing 100190, China}

	\author{Liang-Zhu Mu}
	\affiliation{School of Physics, Peking University, Beijing 100871, China}
	
	\author{Heng Fan}
	\email{hfan@iphy.ac.cn}
	\affiliation{Institute of Physics, Chinese Academy of Sciences, Beijing 100190, China}
	\affiliation{School of Physical Sciences, University of Chinese Academy of Sciences, Beijing 100190, China}
	\affiliation{CAS Center for Excellence in Topological Quantum Computation, University of Chinese Academy of Sciences, Beijing 100190, China}
	
	\begin{abstract}
		A medium-scale quantum computer with full universal quantum computing capability
		is necessary for various practical aims and testing applications.
		Here we report a 34-qubit quantum virtual machine (QtVM)
		based on a medium server. Our QtVM can run quantum assembly language with
graphic interfaces. The QtVM is implemented with single qubit rotation gate,
single to multiple controlled NOT gates to realize the universal quantum computation.
Remarkably, it can realize a series of basic functions, such as, the ``if'' conditional programming language
 based on single-shot projective measurement results,
 ``for'' iteration programming language, build in arithmetic calculation. The measurement can be single-shot and arbitrary number of multi-shot
 types. In addition, there is in principle no limitation on number of logic gates implemented for quantum computation.
By using QtVM, we demonstrate the simulation of dynamical quantum phase transition of transverse field Ising model by quantum circuits,
where 34 qubits with one million gates are realized.
We also show the realization of programmable Shor algorithm for factoring 15 and 35.
	\end{abstract}
	\pacs{}
	\maketitle
	
	In the past years, great progress has been made in quantum computation and
	quantum simulation \cite{IBM,Martinis-Nature15,Bernien,Zhang,Blatt16,Martinis16,zheng17,song17,Xu}.
	The present universal quantum computers based on physical qubits are around dozens of qubits.
	Still, its operation is limited by the number of logic gates and the fidelity is not satisfying.
	The implementation of a quantum computer with relatively large number of qubits and full universal computing capability
	is still challenging.
	On the other hand, we expect that a universal quantum computer will be powerful in solving some specific
	problems with quantum speedup.
	In particular, much work has been done to transfer various problems
	to quantum circuits which can be processed by a quantum computer. For instance,
	the software package OpenFermion enables the simulation of fermionic models and quantum chemistry
	models on quantum hardware with outputs in a string a quantum gates \cite{OpenFermion}.
A platform possessing either physical or digital qubits for universal quantum computing is in demanding.

Efforts have also been made to simulate quantum circuits by classical computer, for example,
to give output by entire final vector state or just provide amplitude for any given qubit-string in
computational bases \cite{classical1,classical2,classical3,classical4,classical5,classical6}. Those two approaches have respectively advantages on depth of the circuits
or number of qubits realizable. In general, there is a trade-off relation for these two advantages,
balancing between either universality about the tasks or the number of qubits.
Here, we report a universal quantum computing simulator, in name of quantum virtual machine (QtVM).
It performs like a real quantum computer. The QtVM is realized on a medium server.
There is no limitation on the depth of the circuits. It possesses
34 qubits for universal quantum computation.

Our QtVM has both graphic interfaces and quantum assembly (QtASM) language for input.
The input file is compatible with QASM such that the QASM codes can directly run
	on QtVM.
	With this platform, we can immediately perform a series of
	quantum computing tasks, for example, simulate physics of quantum many-body systems
	with high precision and run programmable quantum algorithms.
	Our QtVM is simple for operation, various measurement results are attainable such as
single-shot, multi-shot, or the entire vector of the output state. 	
With this platform online accessible for public, it is then in demanding
	to design more algorithms and explore more practical applications for such kind of quantum computers.
	
\begin{figure}
			\includegraphics[width=0.4\textwidth]{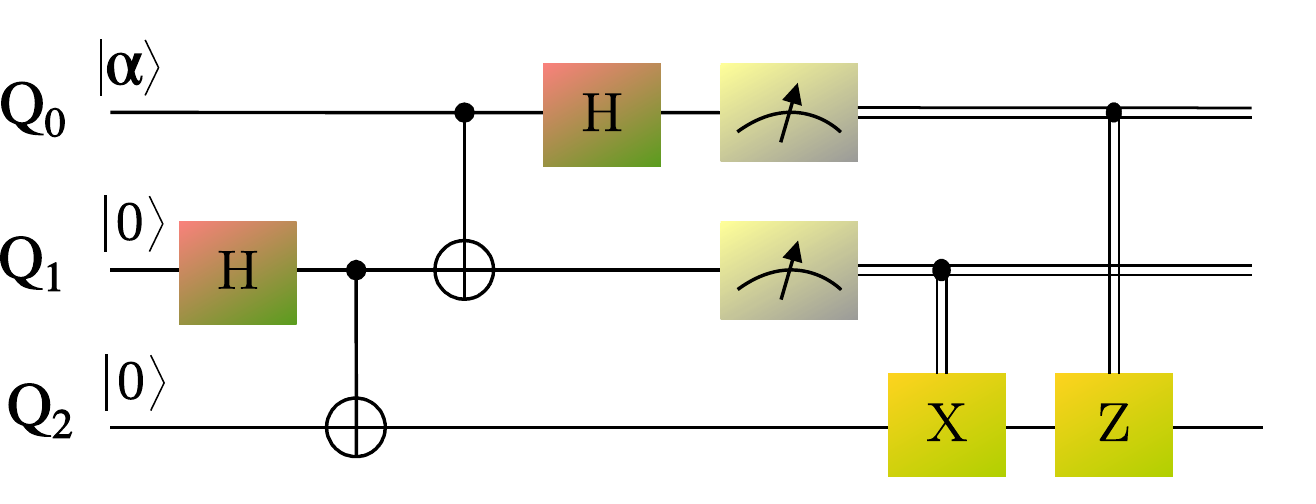}
			\caption{Circuit for teleportation.
This circuit includes quantum operations depending on the
single-shot measurement results. The three qubits are labeled as 0,1,2 from up to down.
The codes to realize this circuit are as follows, $h(1)$, $cnot(1,2)$,
$cnot(0,1)$, $h(0)$, $meas(0,1)$ ,
$meas(1,2)$, $cif(2,`x(2)'), cif(1,`z(2)')$. Code $meas(0,1)$ implies to measure qubit 0 and
put the result to classical register labeled as 1. Code $cif$ means that if the value in classical register is 1, perform the quantum operation; if it is 0, do nothing.
 }
 \label{teleportation}
 \end{figure}

The QtVM is defined as a computing machine, its behaviour is like a \textit{L}-qubit register and a classic register.
It is capable of realizing quantum logic gates and performing projective measurements in computational basis on the quantum register.
The single-shot measuring results are put into classic register.
The arithmetical ability on the classic register and jumping for quantum operation is implemented.
This basic function is necessary, for example, in implementing the logical gates teleportation scheme, see Fig.~\ref{teleportation}
for explanations and codes.

Operations on the QtVM is defined by its instruction set.
It is consisted of quantum gates ($X$,$Y$,$Z$,$S$,$S^{\dagger}$,$H$,C-NOT),
arbitrary single-qubit gates, arbitrary multi-qubits control-unitary gates and classic instructions including measuring,
conditional jump and unconditional jump.
The quantum instructions are closely bounded up with quantum operations on real device and its abstraction capabilities of representing arbitrary gates is very useful for program optimizing and portable binary distribution.
	In addition, QtVM has a friendly interface which is defined by a qtvm\_context C structure, therefore, it enables to program and organize a series of QtVM instance with object oriented programming (OOP) model.
	
It is known that the classic memory required to store the state of a \textit{L}-qubit quantum register grows exponentially with respect to \textit{L} and hard to be stored in the memory. Meanwhile, the storage used to store the matrix representation of a quantum operation also grows exponentially, even making used of sparse matrix formats such as compressed sparse row (CSR). However, most of the operations on a quantum computer has a high instruct symmetry, which could be represented by product of only several small matrices. Therefore, we implemented a basic linear algebra system (BLAS) optimized for those simulation applications.
	The operation is represented by a sequence of QtVM instructions, and some QtVM implementation perform those instructions on data structures, which depends on the system configuration and the size of quantum register.
	
	The lowest level of QtVM simulation implementation is QtVM sector. It operates on two chunks of continuous memory. The real and imaginary parts are stored separately in these chunks so that adjacent positions in the memory store same type of elements, in order to make full use of superscalar instructions. This QtVM implementation provides support of parallel computing on symmetrical multi-processing (SMP).
	
	The second level of QtVM simulation implementation is QtVM pagetable which is designed to handle large-scale simulation efficiently. It manages several QtVM sector instances which each corresponds to a region in the address space of the entire vector. A QtVM lazy-eval instruction queue is assigned to each of the QtVM sector instances. The instructions which could be executed independently on each QtVM sector instance would be pushed into the queue, and the instructions which require interactions between several sectors would either be executed by a quick virtual address reorder or triggering a synchronizing event. When a synchronizing event is triggered, the sequence in each instruction queue will be optimized and then executed by the corresponding QtVM sector. For large scale simulation, the sectors could be configured to be stored in the disk, and the lazy-eval mechanism and preloading subroutine merges I/O requests and reduce page faults.
	It is possible to use the pagetable structure and lazy-eval mechanism to build distributed simulators.
	
	\begin{figure}
		\centering
		\includegraphics[width=0.48\textwidth]{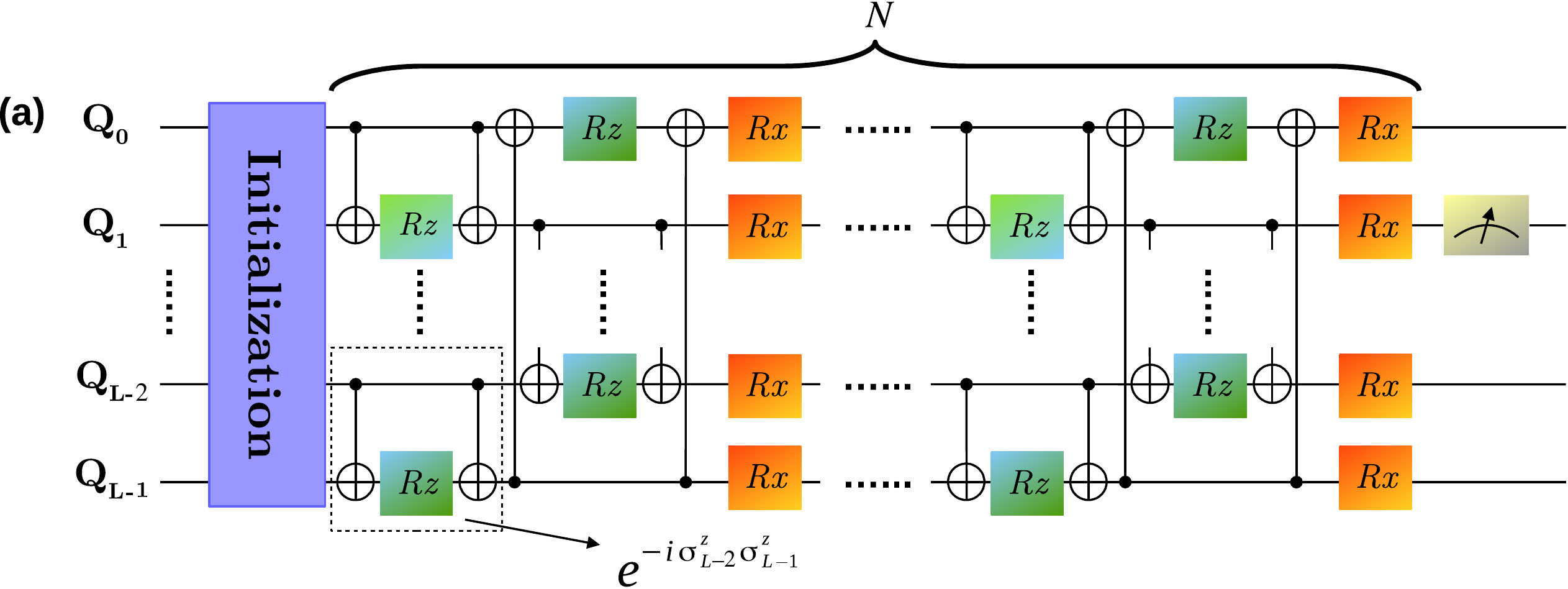}
\vspace{0.25cm}
		\includegraphics[width=0.48\textwidth]{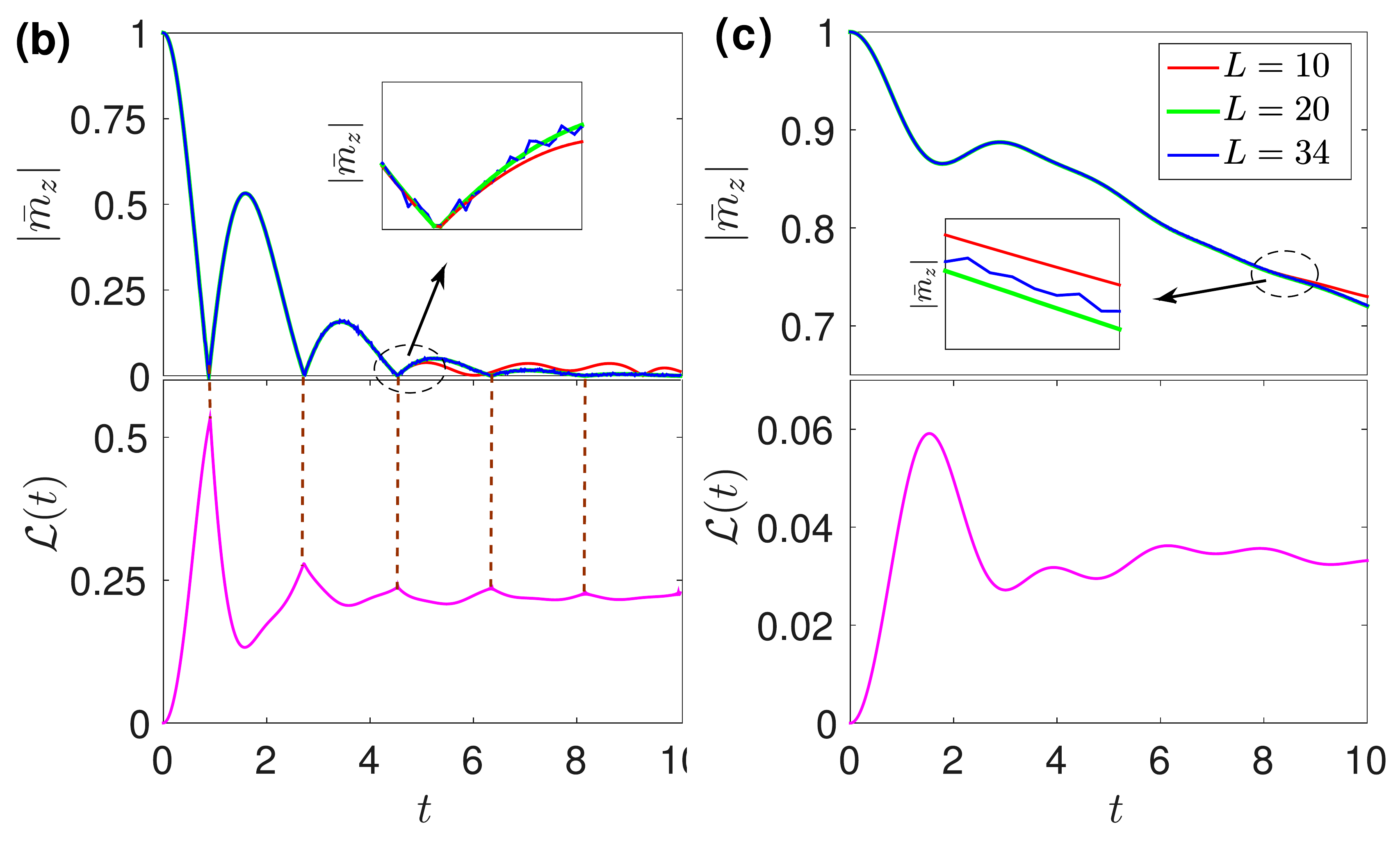}
		\caption{Quantum circuit and simulation. (a) A $L$-qubit quantum circuit for time evolution
of transverse field Ising model is shown. One Suzuki-Trotter step is presented in the beginning (left), and one
 step is presented in the final stage (right). After evolution, a measurement is performed on qubit 1 for average magnetization
 due to translation invariant symmetry.
 We fix $N=500$ steps of Suzuki-Trotter decomposition. Each step is almost the same except the
rotation angles in single qubit gates.  Three cases of qubit
numbers are implemented in QtVM, $L=10,20,34$.
 (b)-(c) The upper panels are dynamics of magnetization after a sudden quench from the same initial condition $H(g_0 = 0)$ to
			$H(g_1 = 2)$ in (b) and $H(g_1 = 0.5)$ in (c), respectively. The lower panels show the function of free energy density.
Here, $\mathcal{L}(t)$ are analytical results in thermodynamical limit. The initial state is $|000...00\rangle$.
In the case of $g_1=2$, when $t = (n+1/2)t^*$ and $t^* \approx 1.83 $,
we find that $\bar{m}_z=0$ correspond exactly to that $\mathcal{L}(t)$ is nonanalytic. The correspondences are shown
by dashed lines in (b). The results demonstrate that the simulation of DQPT
agree well with theoretical expectation. While in case of $g_1=0.5$, $\bar{m}_z$ is
			always larger than $0$ and $\mathcal{L}(t)$ is always analytical, implying no DQPT will occur.}
		\label{dqpt}
	\end{figure}
	
	The consistent intermediate representation and virtual machine (VM) model provides possibility to build universal tools for quantum computing. QtVM currently features an assembly language (ASM) compiler with m4 preprocessor and a C/C++ runtime library, which both have Turing-complete ability. It is also convenient to use external C/C++ scientific library using the runtime library and bind it to other languages. Code optimization routines are provided, which is used in the compilers and QtVM pagetable implementation.
	A debugging tool is also provided, which is able to transverse the probability tree of a quantum program, and inspect the state vector and current instruction sequence.
The time behavior and codes for QtVM are presented in supplemental material.
	
\emph{Simulating quantum dynamics on QtVM.}---One of the main aims of the quantum computer is to simulate properties of many-body systems,
specifically dynamical features \cite{Bernien,Zhang,Xu,Jordens,Schneider,Jotzu,Wu,Atala,Lohse,Nakajima,Flaschner,Greiner,Bloch,Ladd,Devoret,Awschalom,Viola}.
The quantum simulation can be performed on various experimental platforms such as superconducting qubits, trapped ions and cold atoms etc.
Also various numerical methods such as quantum Monte Carlo \cite{Foulkes},
density matrix renormalization group \cite{White,Schollwock} and tensor networks are also successful in investigating various physical problems.

A universal quantum computer can simulate quantum systems by realizing quantum circuits consisting of quantum logical gates.
Here, we present QtVM in emulating dynamical quantum phase transition (DQPT) of Ising model with transverse field \cite{Heyl,Jurcevic}.
The Hamiltonian of the model is written as
	\begin{eqnarray} \label{Hising}
	H(g) = -\sum_{j}\sigma_j^z\sigma_{j+1}^z+g\sigma_j^x,
	\end{eqnarray}
where the periodic boundary condition is assumed.
The time-evolution operator is $U(t)=e^{-itH(g)}$. By applying Suzuki-Trotter decomposition,
	\begin{eqnarray} \label{std}
	e^{\hat{A}+\hat{B}}=\lim_{N\!\rightarrow \infty}(e^{\hat{A}/N}e^{\hat{B}/N})^N
	\end{eqnarray}
with error in order of $O(1/N^2)$, the time-evolution operation can be represented as,
	\begin{eqnarray} \label{ut}
	U(t)\! = \!\big[\!\prod_j\!e^{it/N\sigma_j^z\sigma_{j+1}^z}\!\prod_j\!e^{igt/N\sigma_j^x}\big]^N\!\! +\! O(1/N^2).
	\end{eqnarray}
We then can decompose $U(t)$ into the combination of elementary quantum logical gates.
	
Explicitly, we have $e^{-i\theta\sigma_1^z\sigma_2^z}=\text{C-NOT}_{12}\cdot \text{Rz}_2(\theta) \cdot \text{C-NOT}_{12}$,
	where $\text{C-NOT}_{12}$ is C-NOT gate with qubits 1 and 2 being controlling and target qubits, respectively.
Gate $\text{Rz}(\theta)=e^{-i\theta/2\sigma_z}$ is single qubit rotation gate around \textit{z}-axis with angle $\theta$.
The initial state can be either $|000...00\rangle$ or $|111...11\rangle$, both are ground states of $H(g=0)$ in ferromagnetic phase.
The dynamics of the system is characterized by time evolution of magnetization.
Due to periodic boundary condition, the qubits chain is translational invariant.
Thus, the magnetization $\bar{m}_z$ can be obtained by measuring just one qubit in $z$-axis.
In simulation, we implement $L$ qubits and $N$ steps of Suzuki-Trotter decomposition in QtVM.
For the largest case, we have $L=34$ qubits, $N=500$ steps.
The total number of gates is $6.8\times 10^4$,
however, this number can be further increased. It is obvious that for this simulation,
the $for$ language is necessary to realize the iteration
for steps in Suzuki-Trotter decomposition.

Fig. \ref{dqpt} represents the quantum circuit and the simulation results.
Firstly, we simulate the sudden quench from $g_0 = 0$ to $g_1 = 2$, i.e.,
from ferromagnetic phase to paramagnetic phase across the critical point of phase transition.
	The time evolution curve of magnetization is displayed in Fig.~\ref{dqpt}(b).
	According to Ref.\cite{Heyl}, when $t = (n+1/2)t^*$,
the average magnetization $\bar{m}_z=0$, DQPT will occur.
	Here, $t = \pi/\epsilon_{k^*}(g_1)$, $\cos k^*=(1+g_0g_1)/(g_0+g_1)$, $\epsilon_k(g) = \sqrt{(g-\cos k)^2+\sin^2k}$.
	In this case, $t^* \approx 1.814 $, indeed the critical time obtained from magnetization in simulation
conforms exactly with analytical results in thermodynamical limit.
	Then, let $g_2 = 0.5$, that is, both initial and quenched Hamiltonian are in the same phase.
	We know that there is no DQPT, which means that $\bar{m}_z$ has no zero point.
The simulation result shown in Fig. \ref{dqpt}(c) agrees with the theoretical expectation.

In Supplemental Material, we present more QtVM simulation results of sudden quench dynamics of transverse field Ising model, of which the number of total gates can approach $10^6$ with 34 qubits.
	
		\begin{figure}
			\centering
			\includegraphics[width=0.5\textwidth]{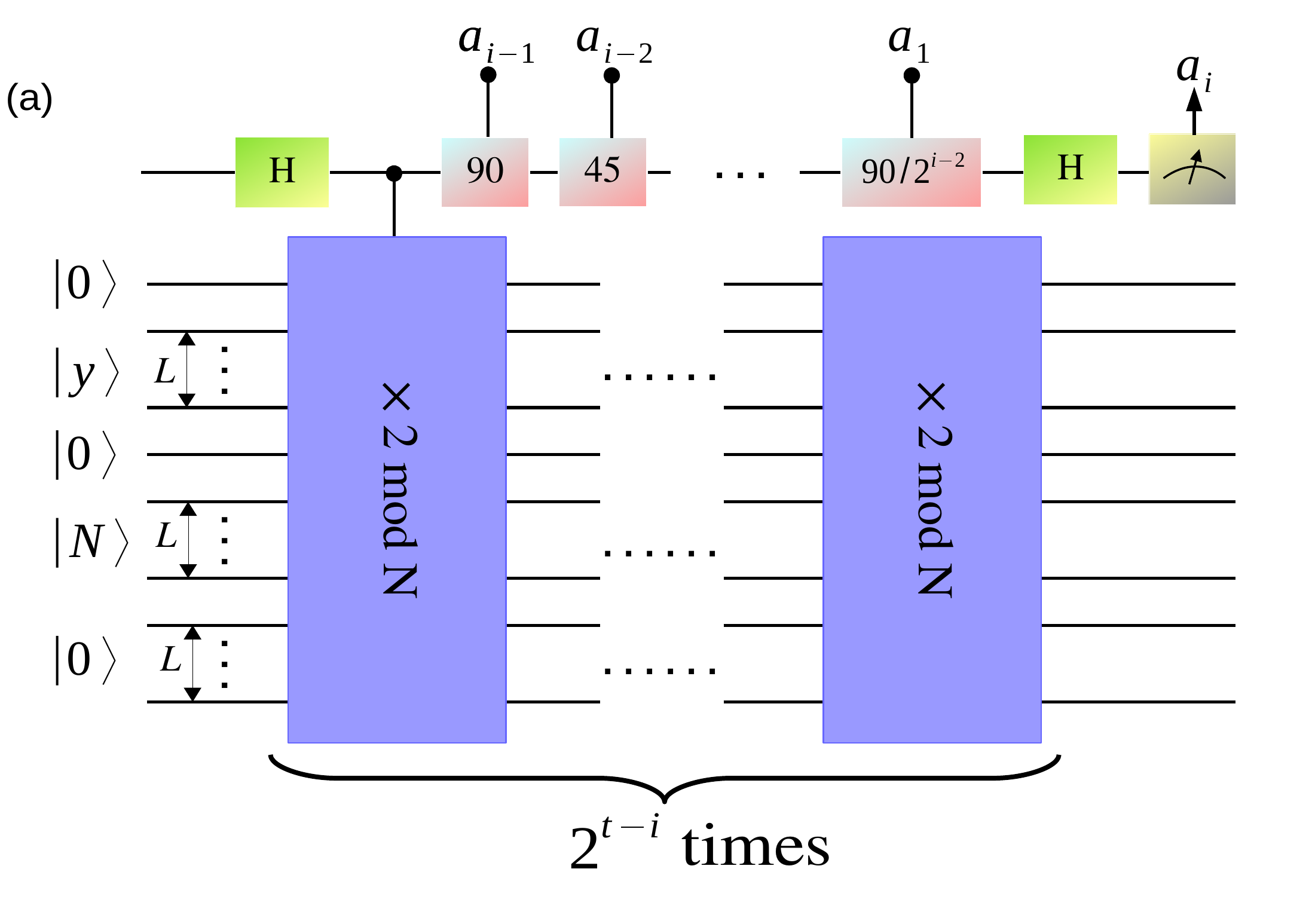}	\vspace{0.25cm}
			\includegraphics[width=0.5\textwidth]{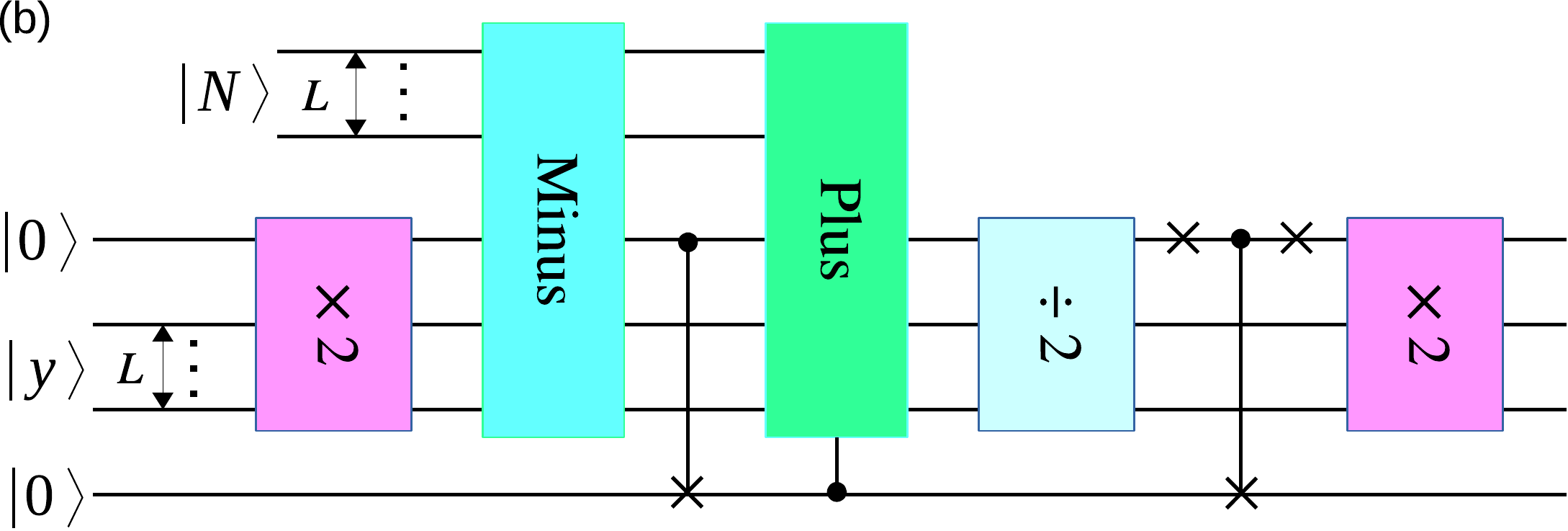}
			\includegraphics[width=0.5\textwidth]{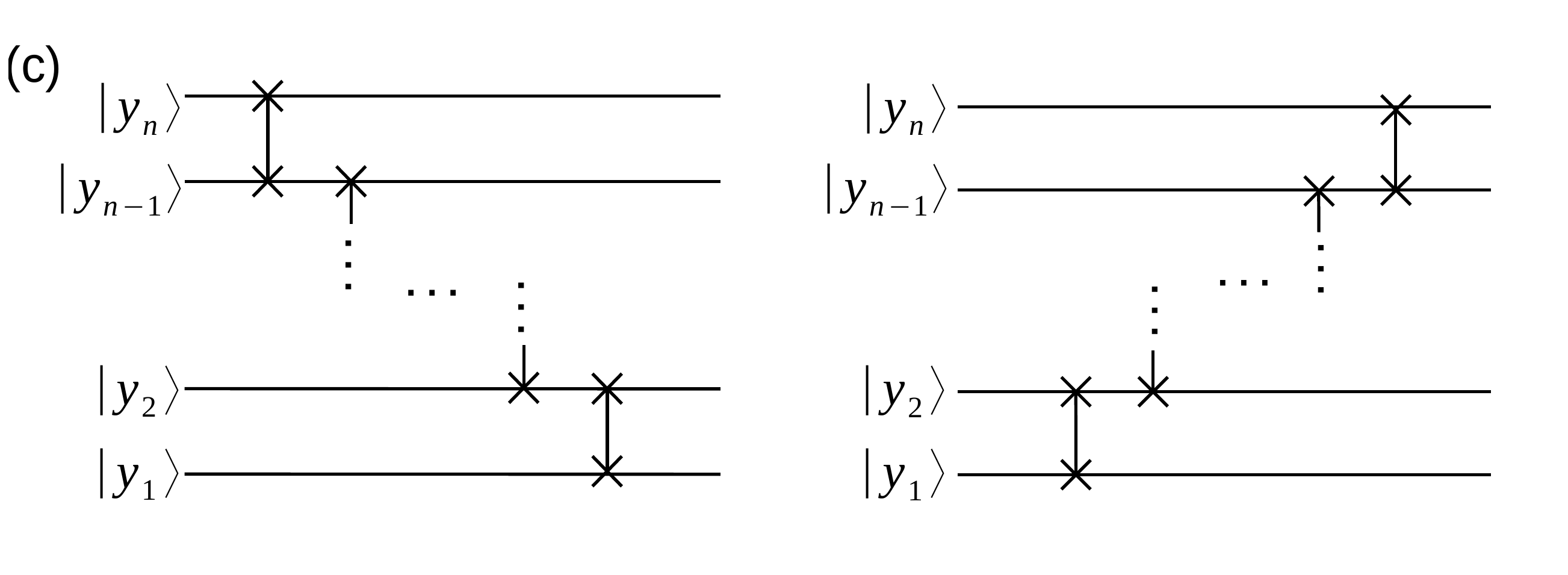}
			\caption{Quantum circuit of Shor algorithm. (a) The quantum circuit of Shor algorithm is presented, where $i$ is an integer among $0,1,...,t-1$ and $L$ is the number of qubits we need to save $N$, satisfying $L=\left\lceil \log2N\right\rceil$. (b)	This figure shows the operation `$\times 2\ \text{mod}\ N$'. To realize the `plus' and `minus' we use the scheme in Ref.\cite{Vedral}, and another $L$ qubits are needed to complete these two modules. (c) The left plot is the operation `$\times 2$', while the right one is `$\div 2$'. They are both the combination of SWAP gates.}
			\label{shor}
		\end{figure}

\emph{Programmable Shor algorithm realized in QtVM.}---		
The main task of quantum computation is to run quantum algorithms \cite{Nielsen}, such as quantum
	Fourier transformation \cite{Coppersmith}, Grover search \cite{Grover} and Shor algorithm
which was realized for special cases in various experimental platforms \cite{Shor,Vandersypen,Politi,Martin,Lucero,Lu,Lanyon}.
Shor algorithm is known for factoring numbers which can be applied to attack the widely used public key cryptosystem.
Now, we use QtVM to demonstrate the realization of Shor algorithm.
The circuits of Shor algorithm are for general purpose, the multiplication of any two
primers can be factored. Thus the codes are applicable for different tasks of factoring.
In this sense, it is programmable. Here we present two cases in factoring 15 and 35 for demonstration.

	Giving two positive coprime integers $x$ and $N$, $x<N$, the order of $x$ modular $N$ is defined as the least positive integer, $r$, such that $x^r=1(\text{mod}\ N)$. The key problem of order-finding is to determine the order for some specified $x$ and $N$.
	When we have an integer $t$ large enough, the order-finding algorithm could solve the problem efficiently.
	The procedure of this algorithm is shown as
	
	\begin{itemize}
		\item $|0\rangle^{\otimes t} |1\rangle $ \\
		\item $ \frac{1}{\sqrt{2^t}}\sum_{j=0}^{2^t-1} |j\rangle |1\rangle $  ( Apply  $ H^{t}$)\\
		\item $\frac{1}{\sqrt{2^t}}\sum_{j=0}^{2^t-1} |j\rangle | j^x \ \text{mod} \ N \rangle $ \\
		$\cong \frac{1}{\sqrt{r2^t}}\sum_{j=0}^{2^t-1} \sum_{s=0}^{r-1} e^{2 \pi  i s j/r} |j\rangle | u_s \rangle $ \\
		(Apply $C-U_{x,N}$) \\
		\item $\frac{1}{\sqrt{N}} \sum_{s=0}^{r-1} |\widetilde{s/r} \rangle | u_s \rangle $  (apply  $F^+$)\\
		\item $\widetilde{s/r} $\\
		\item $r$	.
	\end{itemize}
	Here, the definition of $U_{x,N}$ and $| u_s \rangle$ is as following
	\begin{eqnarray} \label{eq:U}
	U_{x,N} : |y\rangle \rightarrow |xy(\text{mod}\ N)\rangle
	\end{eqnarray}
	\begin{eqnarray}\label{eq:us}
	|u_s\rangle \equiv \frac{1}{\sqrt{r}} \sum_{k=0}^{r-1} \exp[ \frac{-2 \pi i s k}{r} ]| x^k \ \text{mod}\ N \rangle
	\end{eqnarray}

	To compress as much as possible the number of qubits we need, we use a $x$ as $2$ so that we do not need too many qubits to save $x$ and operate the multiplication. We only need swap gates and some C-NOT gates to complete the modular multiplication.
	The whole circuit is composed of $t$ steps. The $i$'th step is shown as Fig.\ref{shor}(a)-(c) ($i$ from $0$ to $i-1$). After each step, we write down the result of the measurement and initialize the first qubit as $|0\rangle$.
	The final result is a bit array as $a_ta_{t-1}...a_1$. And the arrays that most probably come out should be those which satisfy
	$0. a_t a_{t-1} ... a_1 \approx k/r$, where $k$ is an integer among $0,1,...,r-1$.

\begin{figure}
	\centering
	\includegraphics[width=0.5\textwidth]{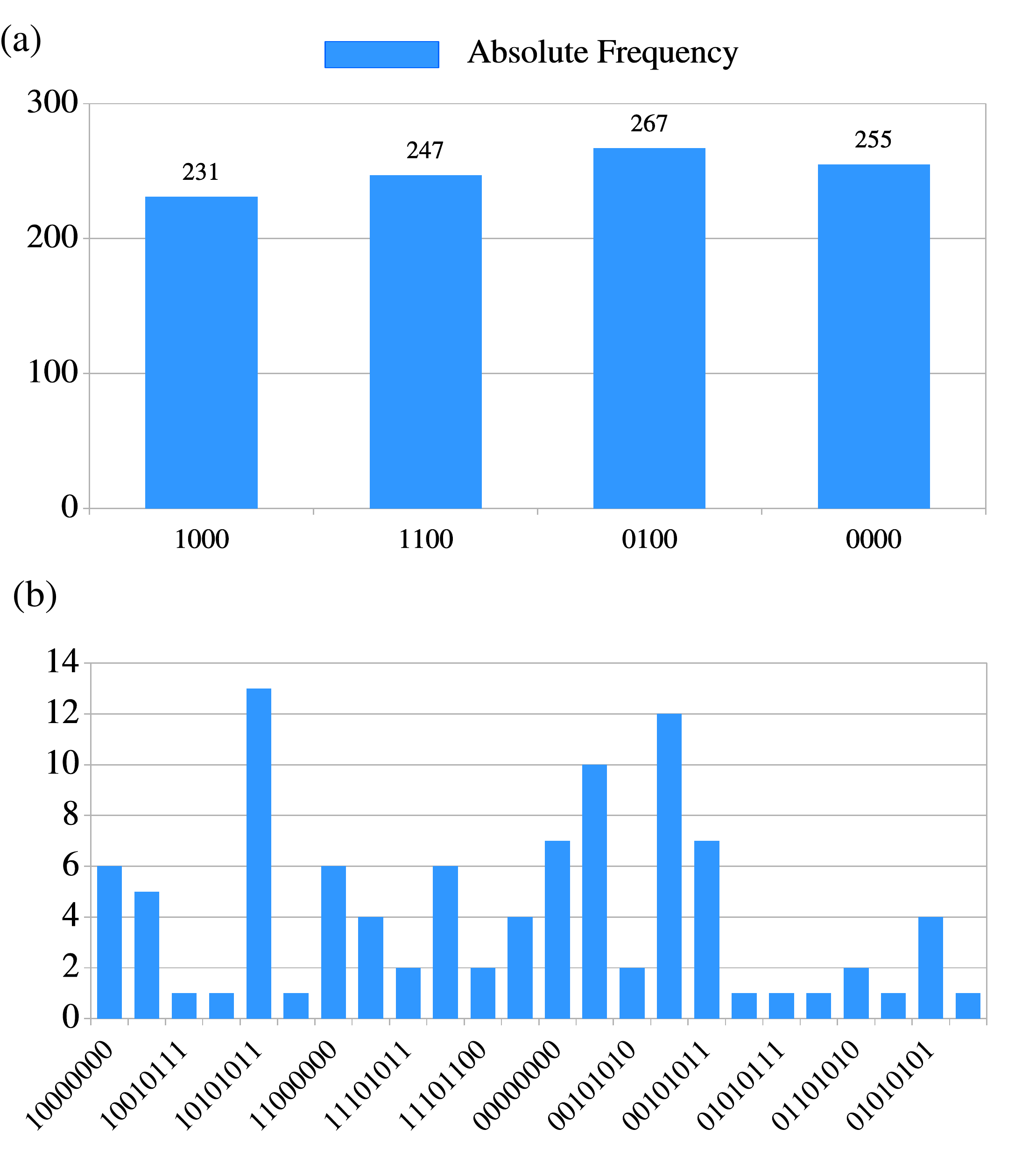}
	\caption{\textbf{(a)} The result of QtVM in factoring $15$. The number of single-shot measurement is 1000.
		The probabilities in four cases are almost equal.
		\textbf{(b)}	The result in factoring $35$.  The number of single-shot measurement is 100. We find that 12 cases happen with or more than 4 times,
		identifying the corresponding output.
	}\label{result_shor}	
\end{figure}

I. {\it Example of factoring 15}. We set state $|N\rangle$ as $|1111\rangle$ which means $N=15$. We use $t$ as $4$ and fetch $4$ bits of measurement results.
For ths case, we require $16$ qubits in QtVM. (The least number of qubit we need is $3L+3$, $L=4$ for factoring 15, $L=6$ for factoring 35).
The procedure repeated for $1000$ times. The result of the procedure is shown as Fig.~\ref{result_shor}(a).

The result $r$ should be $4$. According to Shor algorithm, the possible results of the final classical array, $a_ta_{t-1}...a_1$, of the quantum procedure are $0000$, $0100$, $1000$, $1100$, and the possibilities of them are equal. Fig.~\ref{result_shor}(a) conforms this expectation showing four equal probabilities for the corresponding classical arrays.
Each of the arrays corresponds to a nonnegative integer value of $k$ smaller than $r$. They indicate the four possible results of $k/r$ --- $0.0000$, $0.0100$, $0.1000$ and $0.1100$. The other twelve
$4$-bit arrays are impossible to come out.
The output demonstrated in QtVM as shown in Fig.~\ref{result_shor}(a) shows that we have utilized the order-finding algorithm to find a proper value of $r$, 4.
According to Shor algorithm, the first factor is $gcd(2^{(r/2)}-1,15)=gcd(3,15)=3$ and the other one is $gcd(2^{(r/2)}+1,35)=gcd(5,35)=5$.

II. {\it Example of factoring 35}. We set state $|N\rangle$ as $|100011\rangle$ to factor $35$, $t=8$ and fetch $8$ bits for output.
We need $22$ qubits, the procedure repeated for $100$ times. One result is shown as Fig.~\ref{result_shor}(b), another one is shown in supplementary material.

Slightly different from the case of factoring $15$, when we factor $35$, the probabilities of the possible results of the final classical arrays are not equal.
Actually if we repeat the procedure for infinite times, the frequency function of the arrays will show $12$ sharp peaks. Each of the peaks represent a binary decimal, $0.a_ta_{t-1}...a_1$, and corresponds to a nonnegative integer $k$ smaller than $r$ and a $k/r$.

In Fig.~\ref{result_shor}(b), the result $r$ should be $12$ and we find $12$ arrays that happen more than $4$ times. They form the $12$ peaks.
Their values of the binary decimals, $0.a_ta_{t-1}...a_1$, are very close to the $k/r$ , showing that we have obtained a proper result using the order-finding algorithm.
According to Shor algorithm, the first factor is $gcd(2^6-1,35)=gcd(63,35)=7$ and the other one is $gcd(2^6+1,35)=gcd(65,35)=5$.

\emph{Conclusion and perspective.}---In summary, QtVM behaves like a universal quantum computer.
It can run quantum algorithms and perform quantum simulation.
We should emphasize that the digital simulator for universal quantum computation meets the exponential problems for such as memory in terms of
qubits number. QtVM is a perfect platform in testing quantum algorithms and for quantum simulation
with relatively large number of qubits.

QtVM is a classical computer. It is useful in exploring the bound of quantum supremacy. It can be used for various quantum computation tasks
based on its universality. It is accessible online \cite{online}.

\begin{acknowledgments}
	Q.T.H. wrote the codes of QtVM, W. W. wrote the codes of Shor algorithm, Z. Y. G. wrote the codes of
 quantum simulation. H. F. proposed the project and debugging the codes of Shor algorithm. All authors tested QtVM and proposed suggestions.
 This work was supported by National Key Research and Development Program of China (Grant Nos. 2016YFA0302104 and 2016YFA0300600),
	National Natural Science Foundation of China (Grant Nos. 91536108, 11774406),
Chinese Academy of Sciences (XDB). We thank Zhi-Huai Chen for writing the register system for us.
\end{acknowledgments}


%

\newpage

\clearpage

\widetext
\section{Supplemental material}
\subsection{Quench dynamics of transverse field Ising model}
\begin{figure}[H]
		\centering
	\includegraphics[width=1\textwidth]{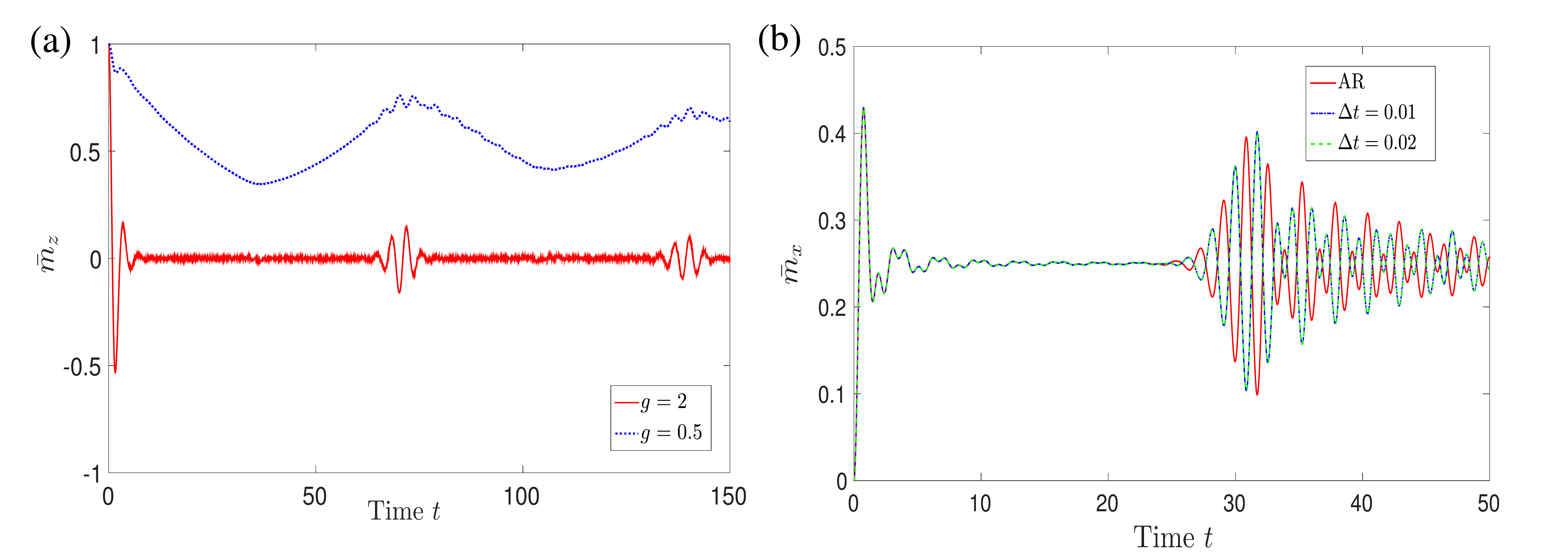}
	\caption{ Simulation results of sudden quench dynamics of transverse field Ising model. \textbf{(a)} The dynamics of magnetic susceptibility in $z$-axis with 34 qubits. Here, $\Delta t = 0.02$, 7500 steps of Suzuki-Trotter decomposition are performed. The total number of logical gates is about
{\bf one million}, that is, $34\times (3+1)\times 7500=1,020,000$. This calculation is performed in about 20 days in QtVM. \textbf{(b)} The dynamics of magnetic susceptibility in $x$-axis with 30 qubits and $g = 2$. }
	\label{qd}	
\end{figure}

Here, we present the simulation results of sudden quench dynamics of transverse field Ising model. Similarly, we choose $|00,...,0\rangle$ as the initial state.  For longitudinal susceptibility, i.e., $\bar{m}_z$, when quenching to the paramagnetic phase, $\bar{m}_z$ tends to zero, while when quenching to ferromagnetic phase, $\bar{m}_z$ tends to a finite value. The explicit simulation results by meas of QtVM are shown in Fig. \ref{qd}(a). For transverse susceptibility, i.e., $\bar{m}_x$, there is a analytical solution
	\begin{eqnarray}\label{as}
   m_x(t) = \frac{2}{L}\sum_{0<k<\pi}\big[\frac{g_0+\cos k}{\omega_0}+\frac{(g-g_0)\sin^2k}{\omega^2\omega_0}(1-4\cos \omega t)\big].
	\end{eqnarray}
Here, $\omega = \epsilon_k(g)$, $\omega_0 = \epsilon_k(g_0)$. See Fig. \ref{qd}(b), the red line is the analytical result, while the other two dashed lines are results of QtVM. We find that, for Suzuki-Trotter decomposition, when $\Delta t = 0.02$, the result can converge for long time. However, since the analytical solution is obtained by Jordan-Wigner transformation, the boundary correction term is neglected. Therefor, the analytical solution and simulation result will exist deviation after long time enough.

\newpage

\clearpage
\begin{figure}
		\centering
		\includegraphics[width=0.6\textwidth]{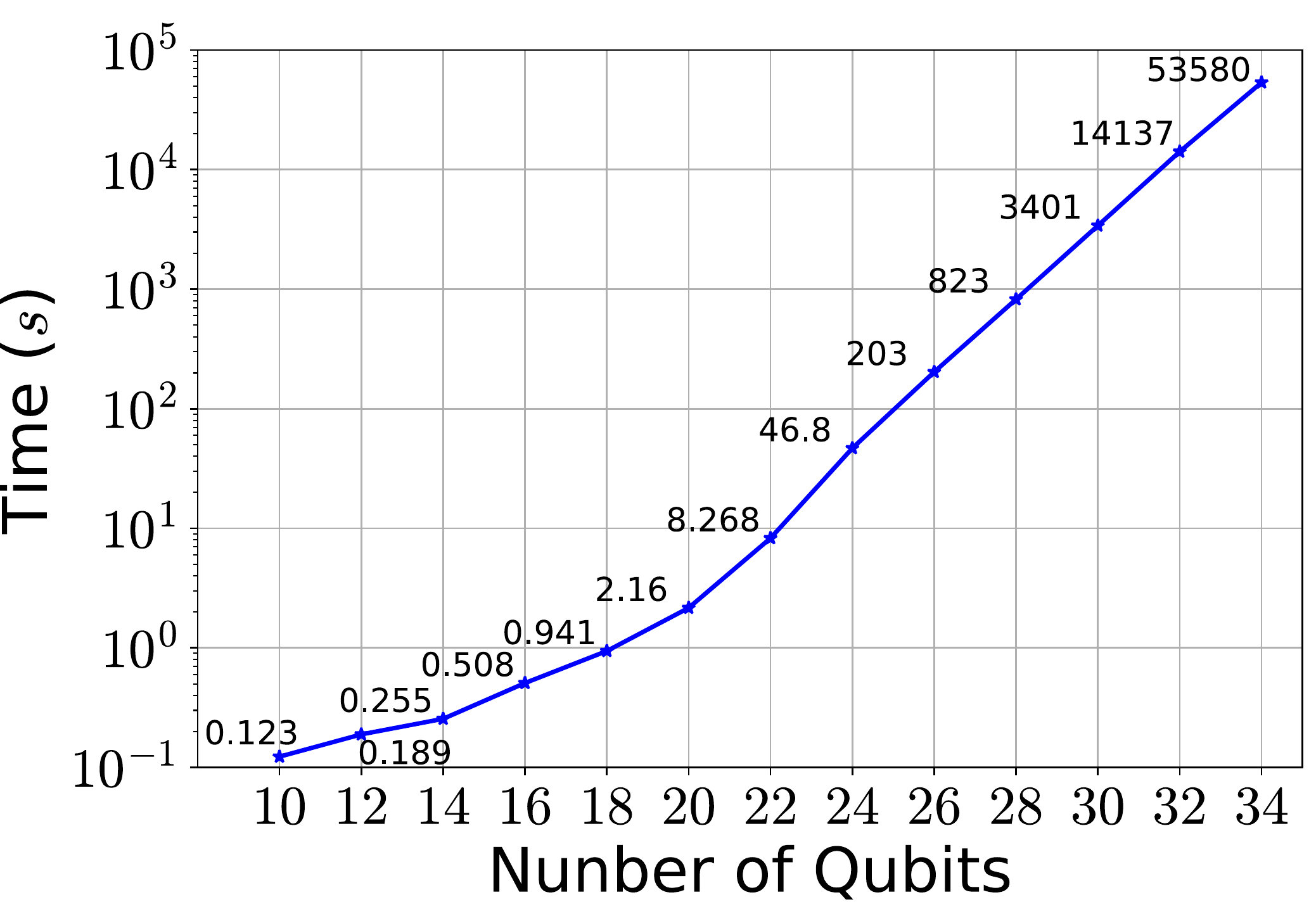}
		\caption{\label{shor2}
			The performance of QtVM. The figure shows the running time (y axis) for 200 gates with 100
single-shot measurements for different number of qubits (x axis). We emphasize that the 100 single-shot
measurements means that we should repeat 100 times the task.
The 200 gates include 100 single qubit gates and
  100 CNOT gates. The running time grows almost exponentially for number of qubits.
		}\label{time}	
	\end{figure}

\begin{figure}
		\centering
		\includegraphics[width=0.6\textwidth]{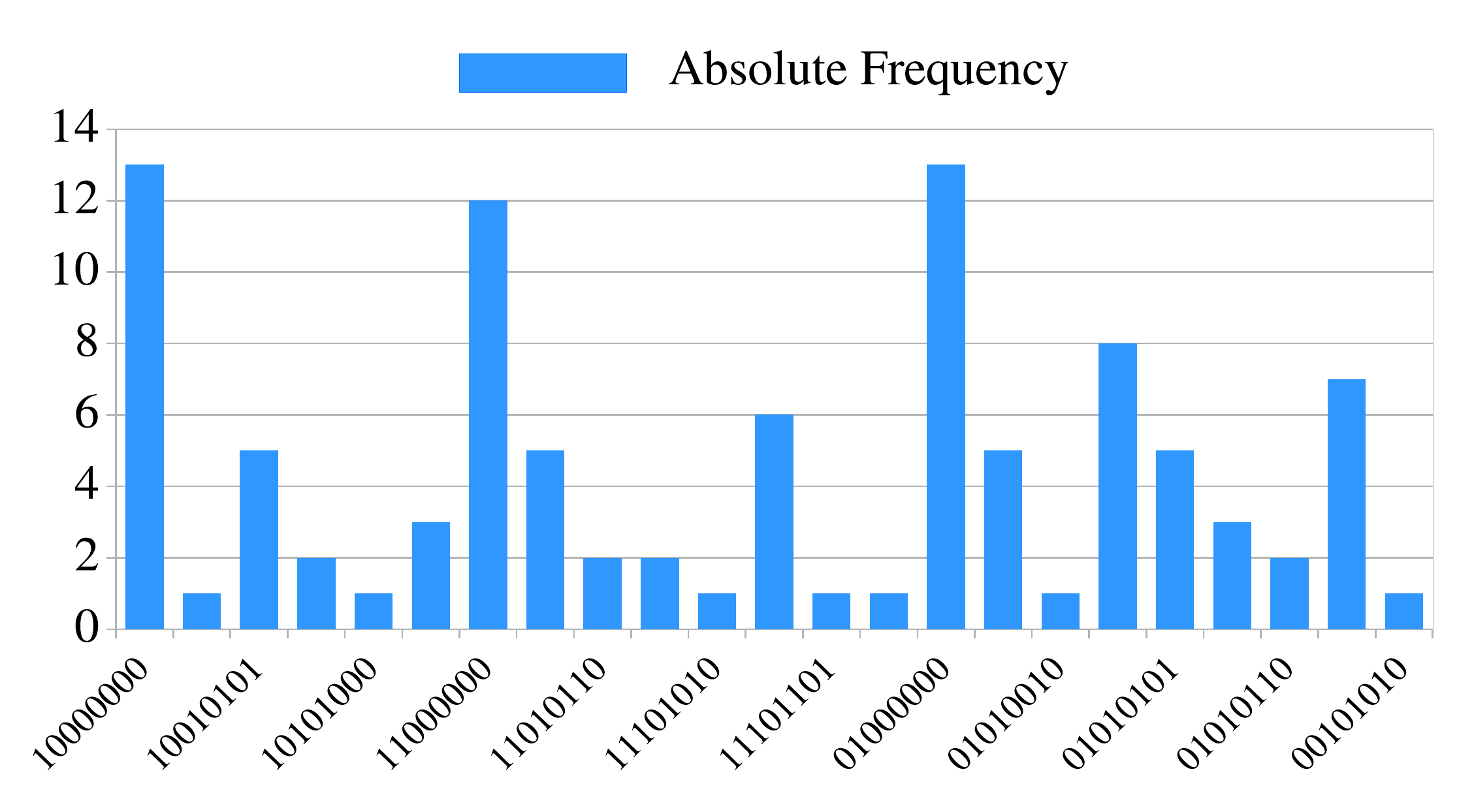}
		\caption{\label{res_shor}
			A different result in factoring $35$. As in the main text, the number of single-shot measurement is 100. We find that 12 cases happen with or more than 3 times
		}\label{result57}	
	\end{figure}

\begin{figure*}[!h]\label{code1}	
			\centering
			\includegraphics[width=1\textwidth]{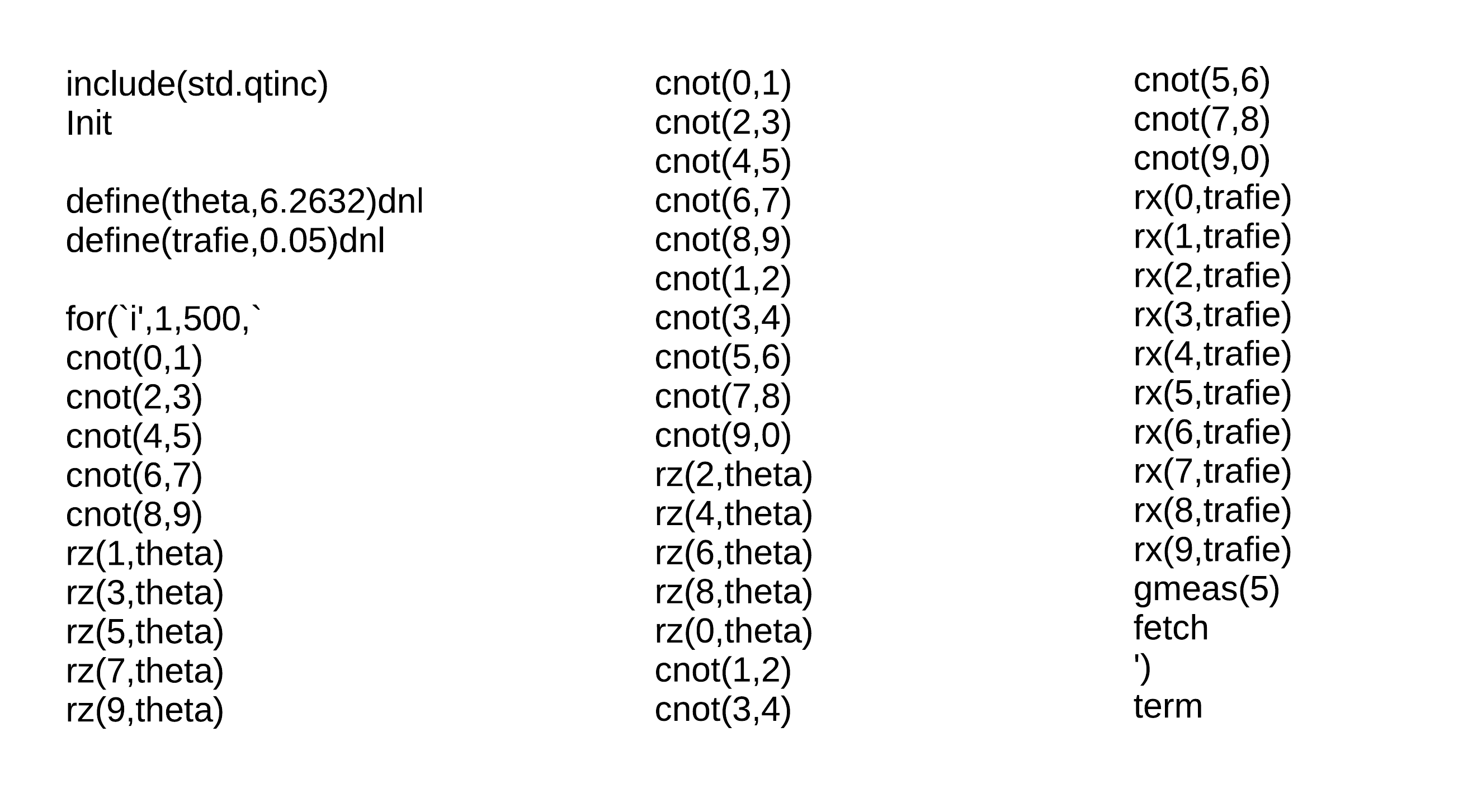}
			\caption{ The QtVM code for simulating dynamical phase transition with 10 qubits.}	
		\end{figure*}

		\begin{figure*}\label{code2}	
			\centering
			\includegraphics[width=1\textwidth]{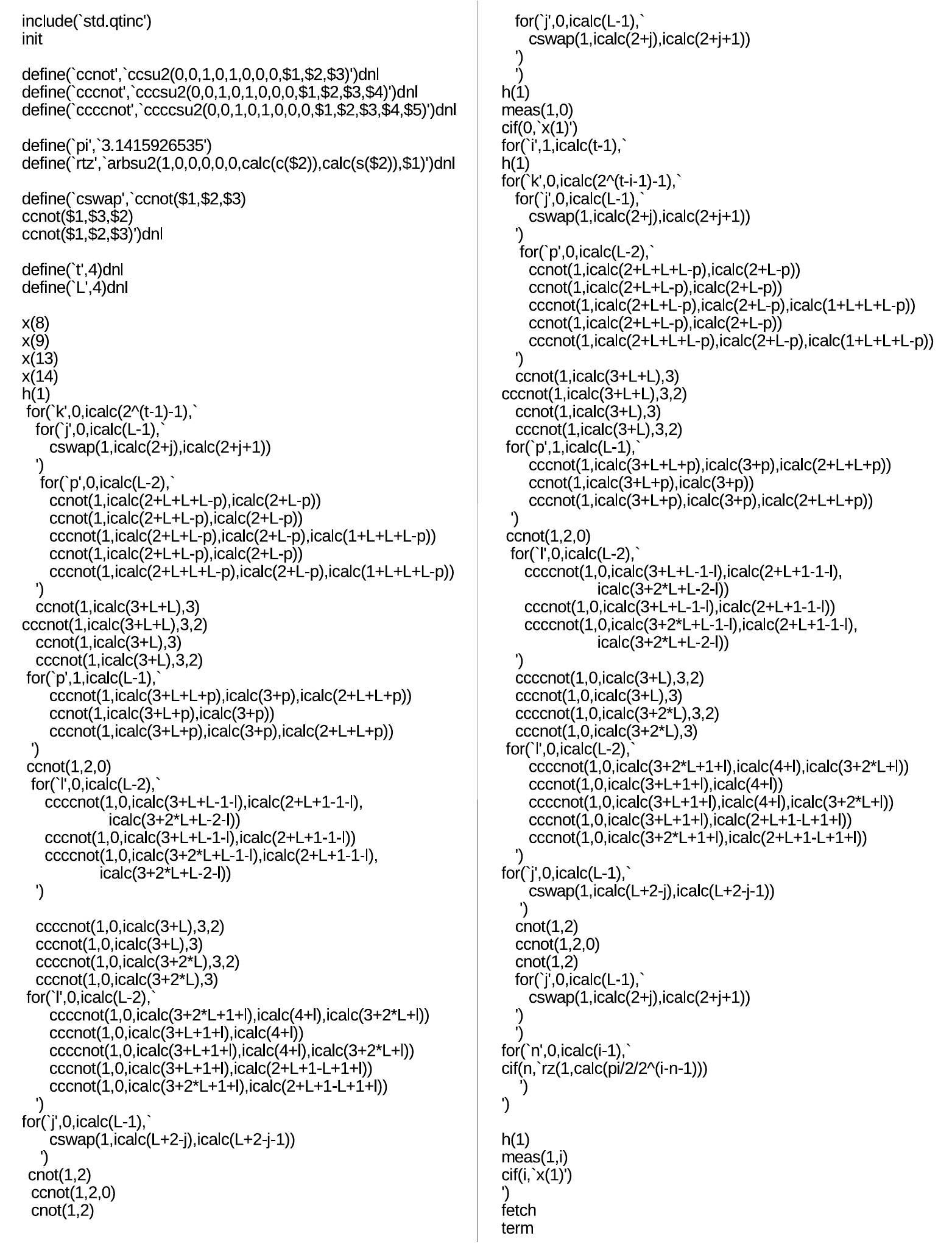}
			\caption{The code for Shor algorithm.}	
		\end{figure*}


\begin{thebibliography}{64}%
\makeatletter
\providecommand \@ifxundefined [1]{%
 \@ifx{#1\undefined}
}%
\providecommand \@ifnum [1]{%
 \ifnum #1\expandafter \@firstoftwo
 \else \expandafter \@secondoftwo
 \fi
}%
\providecommand \@ifx [1]{%
 \ifx #1\expandafter \@firstoftwo
 \else \expandafter \@secondoftwo
 \fi
}%
\providecommand \natexlab [1]{#1}%
\providecommand \enquote  [1]{``#1''}%
\providecommand \bibnamefont  [1]{#1}%
\providecommand \bibfnamefont [1]{#1}%
\providecommand \citenamefont [1]{#1}%
\providecommand \href@noop [0]{\@secondoftwo}%
\providecommand \href [0]{\begingroup \@sanitize@url \@href}%
\providecommand \@href[1]{\@@startlink{#1}\@@href}%
\providecommand \@@href[1]{\endgroup#1\@@endlink}%
\providecommand \@sanitize@url [0]{\catcode `\\12\catcode `\$12\catcode
  `\&12\catcode `\#12\catcode `\^12\catcode `\_12\catcode `\%12\relax}%
\providecommand \@@startlink[1]{}%
\providecommand \@@endlink[0]{}%
\providecommand \url  [0]{\begingroup\@sanitize@url \@url }%
\providecommand \@url [1]{\endgroup\@href {#1}{\urlprefix }}%
\providecommand \urlprefix  [0]{URL }%
\providecommand \Eprint [0]{\href }%
\providecommand \doibase [0]{http://dx.doi.org/}%
\providecommand \selectlanguage [0]{\@gobble}%
\providecommand \bibinfo  [0]{\@secondoftwo}%
\providecommand \bibfield  [0]{\@secondoftwo}%
\providecommand \translation [1]{[#1]}%
\providecommand \BibitemOpen [0]{}%
\providecommand \bibitemStop [0]{}%
\providecommand \bibitemNoStop [0]{.\EOS\space}%
\providecommand \EOS [0]{\spacefactor3000\relax}%
\providecommand \BibitemShut  [1]{\csname bibitem#1\endcsname}%
\let\auto@bib@innerbib\@empty


\bibitem{IBM}IBM Q, \href{http://www.research.ibm.com/ibm-q/}{http://www.research.ibm.com/ibm-q/}.

\bibitem{Martinis-Nature15}J. Kelly, R. Barends, A. G. Fowler, and A. Megrant \textit{et al.},
\textit{State preservation by repetitive error detection in a superconducting quantum circuit}.
\href{https://www.nature.com/articles/nature14270}{Nature (London) {\bf 66}, 519 (2015)}.


\bibitem{Blatt16}T. Monz, D. Nigg, E. A. Martinez, M. F. Brandl,
P. Schindler, R. Rines, S. X. Wang, I. L. Chuang, and
R. Blatt,
\textit{Realization of a scalable Shor algorithm},
\href{http://science.sciencemag.org/content/351/6277/1068}{Science {\bf 351}, 1068 (2016)}.

\bibitem{Martinis16}R. Barends, A. Shabani, L. Lamata, J. Kelly, A. Mezzacapo, and U. Las Heras \textit{et al.},
\textit{Digitized adiabatic quantum computing with a superconducting circuit},
\href{https://www.nature.com/articles/nature17658}{Nature (London) {\bf 534}, 222 (2016)}.


\bibitem{zheng17} Yarui Zheng, \textit{et al.}, \textit{Solving Systems of Linear Equations with a Superconducting Quantum Processor},
\href{http://journals.aps.org/prl/abstract/10.1103/PhysRevLett.118.210504}{Phys. Rev. Lett \textbf{118}, 210504 (2017)}.

\bibitem{song17} Chao Song, \textit{et al.}, \textit{10-qubit entanglement and parallel logic operations with a superconducting circuit},
\href{http://journals.aps.org/prl/abstract/10.1103/PhysRevLett.119.180511}{Phys. Rev. Lett \textbf{119}, 180511 (2017)}.

\bibitem{Bernien}
H. Bernien, S. Schwartz, A. Keesling, H. Levine, A. Omran,
H. Pichler, S. Choi, A. S. Zibrov, M. Endres,
M. Greiner, V. Vuleti'c, and M. D. Lukin,
\textit{Probing many-body dynamics on a 51-atom quantum simulator},
\href{https://www.nature.com/articles/nature24622}{Nature (London) {\bf 551}, 579 (2017)}.

\bibitem{Zhang}J. Zhang, G. Pagano, P. W. Hess, A. Kyprianidis,
P. Becker, H. Kaplan, A. V. Gorshkov, Z.-X. Gong, and
C. Monroe,
\textit{Observation of a Many-Body Dynamical Phase Transition with a 53-Qubit Quantum Simulator},
\href{https://www.nature.com/articles/nature24654}{Nature (London) {\bf 551}, 601 (2017)}.


\bibitem{Xu} K. Xu, \textit{et al.}, \textit{Emulating many-body localization with a superconducting quantum processor},
\href{http://journals.aps.org/prl/abstract/10.1103/PhysRevLett.120.050507}{Phys. Rev. Lett \textbf{120}, 050507 (2018)}.

\bibitem{OpenFermion}Jarrod R. McClean, Ian D. Kivlichan, Damian S. Steiger, Kevin J. Sung, and Yudong
Cao \textit{et al.},\textit{ OpenFermion: The Electronic Structure Package for Quantum Computers},
\href{https://arxiv.org/abs/1710.07629}{arXiv:1710.07629}, software package at \href{www.openfermion.org}{www.openfermion.org}.


\bibitem{classical1}T. H\"aner and D. S. Steiger, in \emph{Proceedings of the International Conference for High Performance Computing, Networking, Storage and Anaysis}, 33 (ACM, 2017).

\bibitem{classical2}M. Smelyanskiy, N. P. D. Sawaya, and A. Aspuru-Guzik,
\textit{qHiPSTER: The quantum high performance Software testing environment},
\href{https://arxiv.org/abs/1601.07195}{ arXiv:1601.07195}.


\bibitem{classical3}E. Pednault, J. A. Gunnels, G. Nannicini, L. Horesh, T. Magerlein, E. Solomonik, and R. Wisnieff,
\textit{Breaking the 49-Qubit barrier in the simulation of quantum circuits},
\href{https://arxiv.org/abs/1710.05867}{arXiv:1710.05867}.

\bibitem{classical4}S. Boixo, S. V. Isakov, V. N. Smelyanskiy, and H. Neven
\textit{Simulation of low-depth quantum circuits as complex undirected graphical models},
\href{https://arxiv.org/abs/1712.05384}{arXiv:1712.05384}.

\bibitem{classical5}Z. Y. Chen, Q. Zhou, C. Xue, X. Yang, G. C. Guo, and G. P. Guo,
\textit{64-Qubit quantum circuit simulation},
 \href{https://arxiv.org/abs/1802.06952}{arXiv:1802.06952}.

\bibitem{classical6}J. Chen. F. Zhang, C. Huang, M. Newman, Y. Shi,
\textit{Classical simulation of intermediate-size quantum circuits},
\href{https://arxiv.org/abs/1805.01450}{arXiv:1805.01450}.


\bibitem{Jordens}R. Jordens, N. Strohmaier, K. G¨¹nter, H. Moritz, and T. Esslinger,
\textit{A Mott insulator of fermionic atoms in an optical lattice},
\href{https://www.nature.com/articles/nature07244}{Nature (London) {\bf 444}, 204 (2008)}.

\bibitem{Schneider}
U. Schneider, L. Hackerm¨¹ller, S. Will, Th. Best, I. Bloch, T. A. Costi, R. W. Helmes, D. Rasch, and A. Rosch,
\textit{Metallic and Insulating Phases of Repulsively Interacting Fermions in a 3D Optical Lattice},
\href{http://science.sciencemag.org/content/322/5907/1520}{Science {\bf 322}, 1520 (2008)}.

\bibitem{Jotzu}G. Jotzu, M. Messer, R. Desbuquois, M. Lebrat, T.
Uehlinger, D. Greif, and T. Esslinger,
\textit{A Mott insulator of fermionic atoms in an optical lattice},
\href{https://www.nature.com/articles/nature13915}{Nature (London) {\bf 515}, 237 (2014)}.

\bibitem{Wu}
Z. Wu, L. Zhang, W. Sun, X.T. Xu, B.Z. Wang, S.C. Ji,
Y. Deng, S. Chen, X.J. Liu, and J.W. Pan,
\textit{Realization of two-dimensional spin-orbit coupling for Bose-Einstein condensates},
\href{http://science.sciencemag.org/content/354/6308/83}{Science {\bf 354}, 83 (2016)}.

\bibitem{Atala}M. Atala, M. Aidelsburger, J. T. Barreiro, D. Abanin, T.
Kitagawa, E. Demler, and I. Bloch,
\textit{Direct measurement of the Zak phase in topological Bloch bands},
\href{https://www.nature.com/articles/nphys2790}{Nat. Phys. \textbf{9}, 795 (2013)}.

\bibitem{Lohse}M. Lohse, C. Schweizer, O. Zilberberg, M. Aidelsburger, and I. Bloch,
\textit {A Thouless quantum pump with ultracold bosonic atoms in an optical superlattice},
\href{https://www.nature.com/articles/nphys3584}{Nat. Phys. \textbf{12}, 350 (2016)}.

\bibitem{Nakajima}S. Nakajima, T. Tomita, S. Taie, T. Ichinose, H. Ozawa, L. Wang, M. Troyer, and Y. Takahashi,
\textit {Topological Thouless pumping of ultracold fermions},
\href{https://www.nature.com/articles/nphys3622}{Nat. Phys. \textbf{12}, 296 (2016)}.

\bibitem{Flaschner}N. Flaschner, D. Vogel, M. Tarnowski, B. S. Rem, D.-S. L¨¹hmann, M. Heyl, J. C. Budich, L. Mathey, K. Sengstock, and C. Weitenberg,
\textit {Observation of dynamical vortices after quenches in a system with topology},
\href{https://www.nature.com/articles/s41567-017-0013-8}{Nat. Phys. \textbf{16}, 265 (2012)}.

\bibitem{Greiner}M. Greiner, O. Mandel, Theodor W. H\"ansch, and I. Bloch,
\textit{Collapse and revival of the matter wave field of a Bose¨CEinstein condensate},
\href{https://www.nature.com/articles/nature00968}{Nature (London) {\bf 419}, 51 (2002)}.

\bibitem{Bloch}I. Bloch, J. Dalibard, and S. Nascimb\'ene,
\textit {Quantum simulations with ultracold quantum gases},
\href{https://www.nature.com/articles/nphys2259}{Nat. Phys. \textbf{8}, 267 (2012)}.

\bibitem{Ladd}T. D. Ladd, F. Jelezko, R. Laflamme, Y. Nakamura, C. Monroe, and J. L. O¡¯Brien,
\textit{Quantum computers},
\href{https://www.nature.com/articles/nature08812}{Nature (London) {\bf 464}, 45 (2010)}.

\bibitem{Devoret}
M. H. Devoret, and R. J. Schoelkopf,
\textit{Superconducting Circuits for Quantum Information: An Outlook},
\href{http://science.sciencemag.org/content/339/6124/1169}{Science \textbf{339}, 1169 (2013)}.

\bibitem{Awschalom}
D. D. Awschalom, L. C. Bassett, A. S. Dzurak, E. L. Hu, and J. R. Petta,
\textit{Quantum spintronics: engineering and manipulating atom-like spins in semiconductors},
\href{http://science.sciencemag.org/content/339/6124/1174}{Science \textbf{339}, 1174 (2013)}.

\bibitem{Viola} L. Viola, E. Knill, and S. Lloyd, \textit{Dynamical Decoupling of Open Quantum Systems},
\href{http://journals.aps.org/prl/abstract/10.1103/PhysRevLett.82.2417}{Phys. Rev. Lett \textbf{82}, 2417 (2018)}.

\bibitem{Foulkes} W. M. C. Foulkes, L. Mitas, R. J. Needs, and G. Rajagopal, \textit{Quantum Monte Carlo simulations of solids},
\href{http://journals.aps.org/prl/abstract/10.1103/RevModPhys.73.33}{Rev. Mod. Phys. \textit{73}, 33 (2001)}.

\bibitem{White} S. R. White, \textit{Density matrix formulation for quantum renormalization groups},
\href{http://journals.aps.org/prl/abstract/10.1103/PhysRevLett.69.2863}{Phys. Rev. Lett \textbf{69}, 2863 (1992)}.

\bibitem{Schollwock}U. Schollwock, \textit{The density-matrix renormalization group},
\href{http://journals.aps.org/rmp/abstract/10.1103/RevModPhys.77.259}{Rev. Mod. Phys. \textit{77}, 259 (2005)}

\bibitem{Heyl} M. Heyl, A. Polkovnikov, and S. Kehrein, \textit{Dynamical quantum phase
	transitions in the transverse-field Ising model},
\href{http://journals.aps.org/prl/abstract/10.1103/PhysRevLett.110.135704}{Phys. Rev. Lett. \textbf{110},
	135704 (2013)}.


\bibitem{Jurcevic} P. Jurcevic, H. Shen, P. Hauke, C. Maier, T. Brydges,
 C. Hempel, B. P. Lanyon, M. Heyl, R. Blatt, and C. F. Roos,
 \textit{Direct observation of dynamical quantum phase
	transitions in an interacting many-body system},
\href{http://journals.aps.org/prl/abstract/10.1103/PhysRevLett.119.080501}{Phys. Rev. Lett. \textbf{119},
	080501 (2017)}.

\bibitem{Nielsen}
M. A. Nielsen, and I. L. Chuang, \textit{Quantum Computation and Quantum Information}
(Cambridge Univ. Press, 2011).

\bibitem{Coppersmith}D. Coppersmith,
\textit {An approximate Fourier transform useful in quantum factoring},
\href{https://arxiv.org/abs/quant-ph/0201067}
  {arXiv:quant-ph/0201067}.

\bibitem{Grover} L. K. Grover,
\textit{Quantum Mechanics Helps in Searching for a Needle in a Haystack},
\href{http://journals.aps.org/prl/abstract/10.1103/PhysRevLett.79.325}{Phys. Rev. Lett. \textbf{79},
	325 (2007)}.

\bibitem{Shor}P. W. Shor,
\textit {Algorithms for quantum computation: discrete logarithms and factoring},
in \textit{Proceedings  of  the  35th  Annual  Symposium  on
	Foundations  of  Computer  Science},
\href{http://ieeexplore.ieee.org/document/365700/?reload=true}{
	IEEE Press, Los Alamitos, CA, 124 (1994)}.

\bibitem{Vandersypen}L. M. K. Vandersypen, M. Steffen, G. Breyta, C. S. Yannoni, M. H. Sherwood, and I. L. Chuang,
\textit{Experimental realization of Shor's quantum factoring algorithm using nuclear magnetic resonance},
\href{https://www.nature.com/articles/414883a}{Nature (London) {\bf 414}, 883 (2010)}.

\bibitem{Politi}
    A. Politi, Jonathan C. F. Matthews, and J. L. O'Brien,
\textit{Shor¡¯s Quantum Factoring Algorithm on a Photonic Chip},
\href{http://science.sciencemag.org/content/325/5945/1221}{Science \textbf{325}, 1221 (2009)}.

\bibitem{Martin}E. Martin-Lopez, A. Laing, T. Lawson, R. Alvarez, X. Q. Zhou, and J. L. O'Brien,
\textit {Experimental realization of Shor's quantum factoring algorithm using qubit recycling},
\href{https://www.nature.com/articles/nphoton.2012.259}{Nat. Photonics. \textbf{6}, 773 (2012)}.

\bibitem{Lucero}E. Lucero, R. Barends, Y. Chen, J. Kelly, M. Mariantoni, A. Megrant, P. O¡¯Malley, D. Sank, A. Vainsencher, J. Wenner, T. White, Y. Yin, A. N. Cleland and J. M. Martinis,
\textit {Computing prime factors with a Josephson phase qubit quantum processor},
\href{https://www.nature.com/articles/nphys2385}{Nat. Phys. \textbf{8}, 719 (2012)}.

\bibitem{Lu} C. Y. Lu, D. E. Browne, T. Yang, and J. W. Pan,
\textit{Demonstration of a Compiled Version of Shor¡¯s Quantum Factoring Algorithm Using Photonic Qubits},
\href{http://journals.aps.org/prl/abstract/10.1103/PhysRevLett.99.250504}{Phys. Rev. Lett. \textbf{99},
	250504 (2007)}.

\bibitem{Lanyon} B. P. Lanyon, T. J. Weinhold, N. K. Langford, M. Barbieri, D. F. V. James, A. Gilchrist, and A. G. White,
\textit{Demonstration of a Compiled Version of Shor¡¯s Quantum Factoring Algorithm Using Photonic Qubits},
\href{http://journals.aps.org/prl/abstract/10.1103/PhysRevLett.99.250505}{Phys. Rev. Lett. \textbf{99},
	250505 (2007)}.

\bibitem{Vedral}V. Vedral, A. Barenco, and A. Ekert, \textit{Quantum networks for elementary arithmetic operations},
\href{http://journals.aps.org/pra/abstract/10.1103/PhysRevA.54.147}{Phys. Rev. A. \textbf{54},
	147 (1996)}.


\bibitem{online}QtVM, \href{http://q.iphy.ac.cn}{http://q.iphy.ac.cn}.

\end{thebibliography}
\end{document}